\documentclass[preprint,showpacs,12pt]{revtex4}
\usepackage{amsmath,amssymb,amsfonts}
\usepackage{graphicx}
\usepackage{hhline}
\usepackage{bm}
\usepackage{amsmath}
\usepackage{epsfig}

\newcommand{\gvec}[1]{\hbox{\boldmath$#1$\unboldmath}}

\begin{document}         
\begin{flushleft}
\end{flushleft}
\title{Electron-neutrino scattering off nuclei from two different theoretical perspectives}
\author {M. Martini$^{1,2}$, N. Jachowicz$^{1}$, M. Ericson$^{3,4}$, V. Pandey$^{1}$, T. Van Cuyck$^{1}$, 
N. Van Dessel$^{1}$}
\affiliation{$^{1}$Department of Physics and Astronomy, Ghent University, Proeftuinstraat 86, B-9000 Gent, Belgium}
\affiliation{$^{2}$ESNT, CEA-Saclay, IRFU, Service de Physique Nucl\'eaire, F-91191 Gif-sur-Yvette Cedex, France}
\affiliation{$^{3}$Universit\'e de Lyon, Univ. Lyon 1,
 CNRS/IN2P3, IPN Lyon, F-69622 Villeurbanne Cedex, France}
\affiliation{$^{4}$Physics Department, Theory Unit, CERN, CH-1211 Geneva, Switzerland}

\begin{abstract} 
We analyze  charged-current electron-neutrino cross sections on Carbon. 
We consider two different theoretical approaches, on one hand the Continuum Random Phase Approximation (CRPA) which allows a description of giant resonances and quasielastic excitations, on the other hand the RPA-based calculations which are able to describe multinucleon emission and coherent and incoherent pion production as well as quasielastic excitations. We compare the two approaches in the genuine quasielastic channel, and find a satisfactory agreement between them at large energies while at low energies the collective giant resonances show up only in the CRPA approach. 
We also compare electron-neutrino cross sections with the corresponding muon-neutrino ones in order to investigate the impact of the different charged-lepton masses. 
Finally, restricting to the RPA-based approach we compare the sum of quasielastic, multinucleon emission, coherent and incoherent one-pion production cross sections (folded with the electron-neutrino T2K flux) with the charged-current inclusive electron-neutrino differential cross sections on Carbon measured by T2K. We find a good agreement with the data. 
The multinucleon component is needed in order to reproduce the T2K electron-neutrino inclusive cross sections.  
\end{abstract}
\pacs{25.30.Pt, 13.15.+g, 24.10.Cn}
\maketitle 
\section{Introduction}
\label{sec_intro} 
Recent years have seen an accumulation of data on muon-neutrino cross sections on nuclei at intermediate energies \cite{AguilarArevalo:2010zc,AguilarArevalo:2009ww,AguilarArevalo:2010cx,Nakajima:2010fp,AguilarArevalo:2010bm,AguilarArevalo:2010xt,Anderson:2011ce,Abe:2013jth,AguilarArevalo:2013hm,Fiorentini:2013ezn,Fields:2013zhk,Acciarri:2014isz,Abe:2014nox,Higuera:2014azj,Eberly:2014mra,Acciarri:2014eit,Walton:2014esl,Aliaga:2015wva,Abe:2015oar}. 
These measurements have revealed interesting features in different reaction channels. 
For example, the charged-current quasielastic (CCQE) measurement performed by MiniBooNE \cite{AguilarArevalo:2010zc} 
has attracted a lot of attention due to its unexpected 
behavior, reproducible with an unphysical value of the axial mass. This axial mass anomaly is now explained by the inclusion of events in which several nucleons are ejected in the CCQE cross section \cite{Martini:2009uj,Martini:2010ex,Amaro:2010sd,Nieves:2011pp,Bodek:2011ps,Martini:2011wp,Nieves:2011yp,Amaro:2011aa,Lalakulich:2012ac,Nieves:2013fr,Martini:2013sha,Martini:2014dqa,Megias:2014qva,Ericson:2015cva}. 
In the one-pion production channel some questions are still open. 
For instance, various theoretical models \cite{Sobczyk:2014xza,Mosel:2015tja} cannot simultaneously reproduce 
the MiniBooNE \cite{AguilarArevalo:2009ww,AguilarArevalo:2010bm} and the MINERvA \cite{Eberly:2014mra} results.

The wealth of experimental and theoretical results on muon-neutrino cross sections
contrasts with the few published results on electron-neutrino cross sections. 
After the inclusive $\nu_e$ CC total cross sections measured by the Gargamelle bubble chamber in 1978 \cite{Blietschau:1977mu}, the first measurement of inclusive $\nu_e$ CC differential cross sections on Carbon was performed by T2K \cite{Abe:2014agb}. Recently the measurement performed by MINERvA of quasielastic and quasielastic-like differential cross sections on Carbon also appeared \cite{Wolcott:2015hda}.  
A precise knowledge of $\nu_\mu$ and $\nu_e$ cross sections is important in connection to the $\nu_\mu \to \nu_e$ oscillation experiments which aim at the 
determination of the neutrino mass hierarchy and the search for CP violation in the lepton sector. A theoretical  comparison of the  $\nu_\mu$ and $\nu_e$ cross sections was performed by Day and McFarland \cite{Day:2012gb} who
 analyzed  the influence of the final lepton-mass difference on the cross sections as a function of the neutrino energy and of $Q^2$. 
Here we study these differences  focusing on the $\nu_\mu$ and $\nu_e$ differential cross sections. 
In a first part we consider the electron-neutrino cross sections on Carbon using two different theoretical models. 
The first one is the one of Martini \textit{et al.} \cite{Martini:2009uj} which is based on nuclear response functions, 
treated in the random phase approximation (RPA) on top of a local relativistic Fermi gas (LRFG) calculation.
It includes the quasielastic cross section, multinucleon emission and coherent and incoherent single pion production. 
The second model is the one of Jachowicz \textit{et al.} \cite{Jachowicz:2002rr} 
which is based on the continuum random phase approximation (CRPA) on top of Hartree-Fock (HF) calculations.  It was originally developed to study electroweak reactions in the giant resonance region and then extended by Pandey \textit{et al.} \cite{Pandey:2013cca,Pandey:2014tza} to the quasielastic regime. 
The common channel where the two approaches can be compared is hence the quasielastic one. 
After a description of the two theoretical models, we confront their results in the quasielastic channel, first for fixed kinematics. 
We also illustrate in both models the differences between  $\nu_\mu$ and $\nu_e$ cross sections. 
The difference between the two approaches manifests itself for small values of the neutrino energy ($E_\nu\simeq$ 200 MeV). 
However when the cross sections are folded with the T2K and the MiniBooNE $\nu_e$ fluxes which are peaked at $E_\nu\simeq$ 500 MeV it turns out 
that the differences are washed out. This justifies the use of the RPA approach of Martini \textit{et al.}, which incorporates, beyond the quasielastic, 
one pion production and multinucleon emission to compare with the inclusive $\nu_e$ CC differential cross sections on Carbon recently measured by T2K \cite{Abe:2014agb}. We postpone the comparison with the very recent MINERvA results \cite{Wolcott:2015hda} to a future paper.

\section{Theoretical models} 
\label{sec_models}
We summarize here the basic ingredients of the two models. 
Both approaches  calculate the polarization propagator $\Pi$ in the random phase approximation (RPA)
which allows the inclusion of long-range nucleon-nucleon correlations. 
This amounts to solving integral equations which have the generic form 
\begin{equation} \label{eq:rpa_generic}
\Pi = \Pi^0 + \Pi^0 \, V \Pi,
\end{equation}
where $\Pi^0$ is the bare polarization propagator and $V$ denotes the effective particle-hole interaction. 
However, the bare polarization propagator and the residual interaction differ in the two approaches.  

For Martini \textit{et al.} \cite{Martini:2009uj} the \textit{bare} polarization propagator is evaluated in momentum space. 
In a finite system it is non-diagonal and writes $\Pi^0(\omega,\bf{q},{\bf{q}}')$.
In order to account for the finite-size effects, it is evaluated in a semi-classical approximation \cite{Chanfray:1988,Alberico:1997jg} 
where it can be cast in the form 
\begin{equation} \label{eq:laktineh}
\Pi^0(\omega,{\bf{q}},{\bf{q'}}) = \int\, d {\bf{r}} \,e^{-i({\bf{q}}-{\bf{q'}}) \cdot {\bf{r}}} \,
\Pi^0\left(\omega,\frac{{\bf{q}}+{\bf{q'}}}{2},{\bf{r}}\right).
\end{equation}
This semiclassical calculation is a Wigner-Kirwood expansion in $\hbar$ restricted to first order. Its validity has been explored by Stroth \textit{et al.} \cite{Stroth:1985}
 and by Chanfray and Schuck \cite{Chanfray:1988} with the conclusion that the semiclassical description agrees with a quantum mechanical calculation but for small momenta ($q<200$ MeV/c for light nuclei). The non-diagonality in the momenta of $\Pi^0(\omega,{\bf{q}},{\bf{q'}})$ shows that this description goes beyond the infinite nuclear matter one. 
In order to obtain the quantity in the integrand of Eq. (\ref{eq:laktineh}), a local density approximation is used which relates it to the relativistic Fermi gas polarization propagator according to
\begin{equation}
\Pi^0\left(\omega,\frac{{\bf{q}}+{\bf{q}}'}{2},{\bf{r}}\right)=
\Pi^0_{k_F(r)}\left(\omega,\frac{{\bf{q}}+{\bf{q}}'}{2}\right). 
\end{equation}
The local Fermi momentum $ k_F(r) $ is related to the experimental nuclear density through~: 
$ k_F(r) = ( 3/2 \, \pi^2 \, \rho(r) )^{1/3} $.  
The density profile of  $^{12}$C  is taken from the 
Sum-of-Gaussians nuclear charge density distribution parameters according to Ref. \cite{De Jager:1987qc}. 
As for the collective RPA effects, the microscopic approach first introduced by Delorme \textit{et al.} 
\cite{Delorme:1980} is used. The method is described in detail in Ref. \cite{Martini:2009uj}. 
In the bubble RPA chain for the spin-isospin channel the possibility of $\Delta$-hole excitation is taken into account. 

In the approach of Jachowicz \textit{et al.} \cite{Jachowicz:2002rr}, the starting point is the continuum Hartree-Fock model which evaluates the bound and the continuum single-nucleon wave functions through the solution of the Schr\"{o}dinger equation with a mean field potential. The bare polarization propagator, in this case the HF one, is then calculated in coordinate space. According to Ref. \cite{Pandey:2014tza}, the maximum multipole number to calculate the multipole operators is set to $J=16$ which turns out to be sufficient to reach the convergence of the cross sections. 
 
The particle-hole residual interaction differs as well in the two approaches. In the Martini \textit{et al.} one, the  parametrization in terms of pion exchange, rho exchange and contact Landau-Migdal parameters is used 
\begin{eqnarray} \label{eq:INTERACTION}
V_{NN} & = & ( f^\prime \, + \, V_\pi \, + \, V_\rho \, + \, V_{g^\prime} )  \,\, 
\gvec{\tau}_1 . \gvec{\tau}_2 \nonumber \\
V_{N \Delta} & = & ( V_\pi \, + \, V_\rho \, + \, V_{g^\prime} )  \,\, 
\gvec{\tau}_1 . \gvec{T}^\dagger_2 \nonumber \\  
V_{\Delta\Delta} & = & ( V_\pi \, + \, V_\rho \, + \, V_{g^\prime} )  \,\, 
\gvec{T}_1 . \gvec{T}^\dagger_2 . 
\end{eqnarray}
For instance, in the $ NN $ case one has :
\begin{eqnarray}\label{eq:INTERACTION_two}
V_\pi & = &  \left(\frac{g_r}{2 M_N}\right)^{2}\, F_\pi^2(Q^2) \,\, \frac{\gvec{q}^2}{\omega^2 - \gvec{q}^2 - m_\pi^2} \,\,
 \gvec{\sigma}_1 . \widehat{q} \,\, \gvec{\sigma}_2 . \widehat{q} \nonumber \\
V_\rho & = &\left(\frac{g_r}{2 M_N}\right)^{2}\,C_\rho\, F_\rho^2(Q^2) \,\, \, \frac{\gvec{q}^2}{\omega^2 - \gvec{q}^2 - m_\rho^2}  \,\,
 \gvec{\sigma}_1 \times \widehat{q} \,\, \gvec{\sigma}_2 \times \widehat{q} 
\nonumber \\
V_{g^\prime} & = &\left(\frac{g_r}{2 M_N}\right)^{2}\, F_\pi^2(Q^2) \,\,\, g^\prime \,\,\, \gvec{\sigma}_1 . \gvec{\sigma}_2, 
\end{eqnarray}
where $g'$ is the Landau-Migdal parameter and $C_\rho=1.5$. Here $ F_\pi(Q^2) =(\Lambda^{2}_{\pi}- m^{2}_{\pi})/(\Lambda^{2}_{\pi} + Q^2)$ and 
$ F_\rho(Q^2) =(\Lambda^{2}_{\rho}- m^{2}_{\rho})/(\Lambda^{2}_{\rho} + Q^2)$ are the pion-nucleon and rho-nucleon form factors, 
with $\Lambda_{\pi}$ = 1 GeV and $\Lambda_{\rho}$ = 1.5 GeV. Their presence suppresses the residual forces at large $Q^2$. 
For the Landau-Migdal parameter $f^\prime$ the value $f^\prime= 0.6 $ is taken. 
As for the spin-isospin  parameters  $g^\prime $ the following values are used: $g'_{NN}=0.7$, $g'_{N\Delta}=g'_{\Delta\Delta}=0.5$.

In the Jachowicz \textit{et al.} model, the zero-range Skyrme effective interaction, in its parametrization called SkE2 \cite{Waroquier:1979zz,Ryckebusch:1988}, is used. The same interaction which enters in the mean field calculation is employed to generate the continuum RPA (CRPA) solution. In this way, the calculation becomes self-consistent with respect to the interaction.  
In Ref. \cite{Pandey:2014tza} this residual Skyrme-type interaction is multiplied by a dipole form factor which controls the influence of the residual interaction at high $Q^2$ values \cite{Polonica:2009}. 
At variance with respect to Ref.\cite{Pandey:2014tza}, here the CRPA results are not folded with a Lorentz function which would spread the strength of giant-resonances without modification of the position of the quasielastic peak.  

Concerning the RPA differences, an important point should be mentioned. 
The possibility of $\Delta$-hole excitation in the RPA chain is included explicitly in the case of Martini \textit{et al.} 
This is reflected in the appearance in certain kinematical regions of a sizable quenching in the RPA results, due to the mixing of nucleon-hole states with $\Delta$-hole ones, the Ericson-Ericson--Lorentz-Lorenz 
(EELL) effect \cite{Ericson:1966fm}. This quenching has been introduced and established in pion scattering \cite{Ericson:1966fm}. 
It has been discussed also in relation with electron \cite{Alberico:1981sz} and neutrino 
\cite{Nieves:2004wx,Martini:2009uj,Martini:2011wp} scattering.

\section{Comparison between theoretical calculations} 
\label{sec_comparison_theo}
\subsection{LRFG+RPA \textit{vs} HF+CRPA}
\label{subsec_RPA_vs_CRPA}
\begin{figure}
\begin{center}
  \includegraphics[width=15cm,height=10cm]{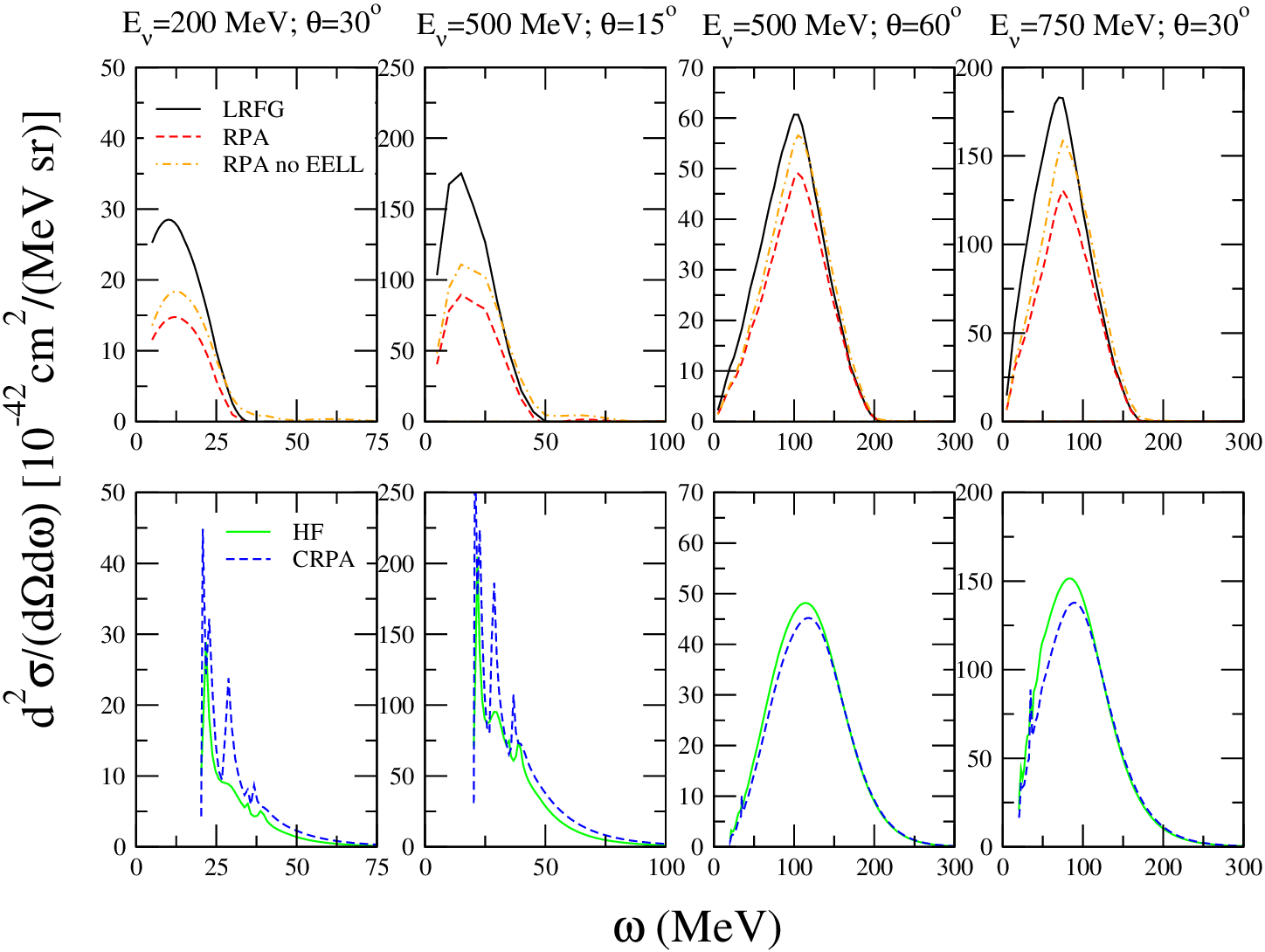}
\caption{(Color online) Electron-neutrino CC double differential cross section on Carbon for fixed values of scattering angles and incident neutrino energies as a function of the energy transferred to the nucleus. In the upper panels the results obtained in the bare-LRFG approach and 
in the RPA with and without the EELL effect are displayed. 
In the lower panels the results obtained in HF and CRPA approaches are displayed.
Only genuine quasielastic and giant resonance excitations (given by the CRPA) are considered.}
\label{fig_LFGRPA_vsMFCRPA}
\end{center}
\end{figure}

\begin{figure}
\begin{center}
  \includegraphics[width=15cm,height=10cm]{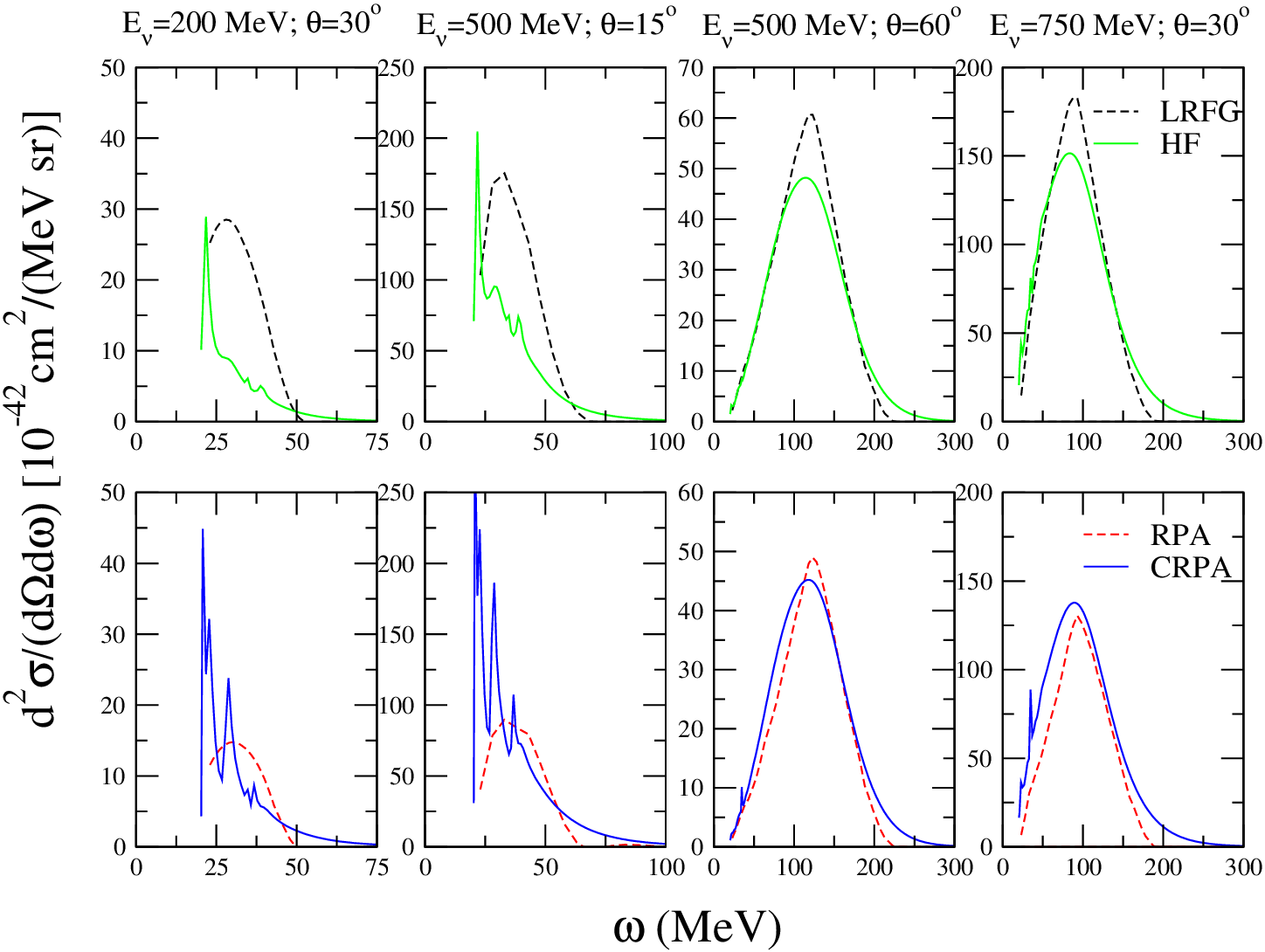}
\caption{(Color online) Electron-neutrino CC double differential cross section on Carbon for fixed values of scattering angles and incident neutrino energies as a function of the energy transferred to the nucleus. 
In the upper (lower) panels the results obtained in the bare-LRFG (RPA) and HF (CRPA) approaches are displayed.
Only genuine quasielastic and giant resonances excitations (given by the CRPA) are considered. Continuous lines: HF and CRPA results; dashed lines: LRFG and RPA results shifted by 18 MeV.}
\label{fig_cfr_d2s_200_500_750_RPA_only}
\end{center}
\end{figure}

In this subsection, we compare the theoretical results in the one nucleon-one hole sector 
obtained in the two different approaches. 
We consider the $\nu_e$-$^{12}$C double differential 
cross sections for different values of the neutrino energy and lepton scattering angle. These cross sections are purely theoretical quantities since 
the experimental  ones depend on the neutrino fluxes and hence are specific for each experiment. 
In Fig.\ref{fig_LFGRPA_vsMFCRPA} we display the results of the two approaches by switching on and off the residual particle-hole interaction. We keep the same notations as in the previous papers of the groups. Namely we call ``bare-LRFG'' the results of Martini \textit{et al.} when the particle-hole interaction is switched off (these are relativistic Fermi gas results in the local density approximation) and ``RPA'' the results obtained by switching on the particle-hole interaction. The corresponding results in the case of Jachowicz \textit{et al.} are called ``HF'' and ``CRPA''. Some important differences between the two approaches appear. 
The most striking feature is the appearance of giant resonance peaks in the CRPA results of Jachowicz \textit{et al.} They vanish for large neutrino energy or larger scattering angle. 
The second comment concerns the threshold energy in the HF+CRPA approach, about $\simeq$ 18 MeV, 
 which reflects the nucleon separation energy, ignored in the semi-classical approximation of Martini \textit{et al.} 
The HF+CRPA results also display the shell structure, which is not present in the semiclassical description. It disappears at large angles or energies, where the two approaches become similar. However, when compared to the semiclassical LRFG results, 
in the mean field HF case the quasielastic peak is somewhat quenched  and the high transferred-energy tail is more important. 
As discussed in Ref. \cite{Stroth:1985} this is a consequence of the non-locality of the mean field which has a similar effect as the residual transverse repulsive ph interaction: it quenches and hardens the responses.  

Turning to RPA effects, the important difference is the large RPA quenching in the Martini \textit{et al.} approach, due to the mixing with $\Delta$-hole states that we have commented before, not explicitly present in the CRPA results of Jachowicz \textit{et al.} For completeness the RPA results without the $\Delta$-hole mixing 
(the EELL effect) are also shown in Fig. \ref{fig_LFGRPA_vsMFCRPA}. In this context further comments are in order. The particle-hole force which enters in the RPA chain, as given in Eqs. (\ref{eq:INTERACTION}), (\ref{eq:INTERACTION_two}), decreases with the momentum transfer. This is apparent in the upper panels 
of Fig. \ref{fig_LFGRPA_vsMFCRPA} from the differences between the bare (LRFG) and RPA cases with or without the EELL effect. Notice also that the influence of the RPA in the approach of Jachowicz \textit{et al.} in the larger momentum cases, even if small, is similar to the one of the Martini \textit{et al.} without the EELL effect: it leads to a small quenching and hardening of the cross section. This behavior is less evident at low momentum transfer, 
where the main difference between HF and CRPA is in the appearance of the giant resonances, however one can notice a hardening of the strength in the CRPA tails.

In order to better illustrate the comparison between the two approaches, we show in Fig.~\ref{fig_cfr_d2s_200_500_750_RPA_only} the LRFG and RPA 
results shifted by 18 MeV, 
an average value of the separation energy, and we compare them with the HF and CRPA results respectively. 
For the structureless part of the cross sections, \textit{i.e.} for 
the kinematical conditions 
such as $\theta=60^o$ and $E_{\nu_e}=500$ MeV or $\theta=30^o$ and $E_{\nu_e}=750$ MeV
, the two approaches are essentially in agreement. Furthermore the HF or CRPA cross sections are characterized by a stronger tail 
at high transferred energies, an effect, as mentioned before, of the non-locality of the mean field. 
It turns out that, in the low-energy part, the RPA results (which do not show giant resonance peaks) represent the average of the CRPA calculations relatively well,  
in spite of the fact that the semi-classical approach is not expected to be valid in this region, which corresponds to low transferred momenta. 

\begin{figure}
\begin{center}
  \includegraphics[width=15cm,height=6cm]{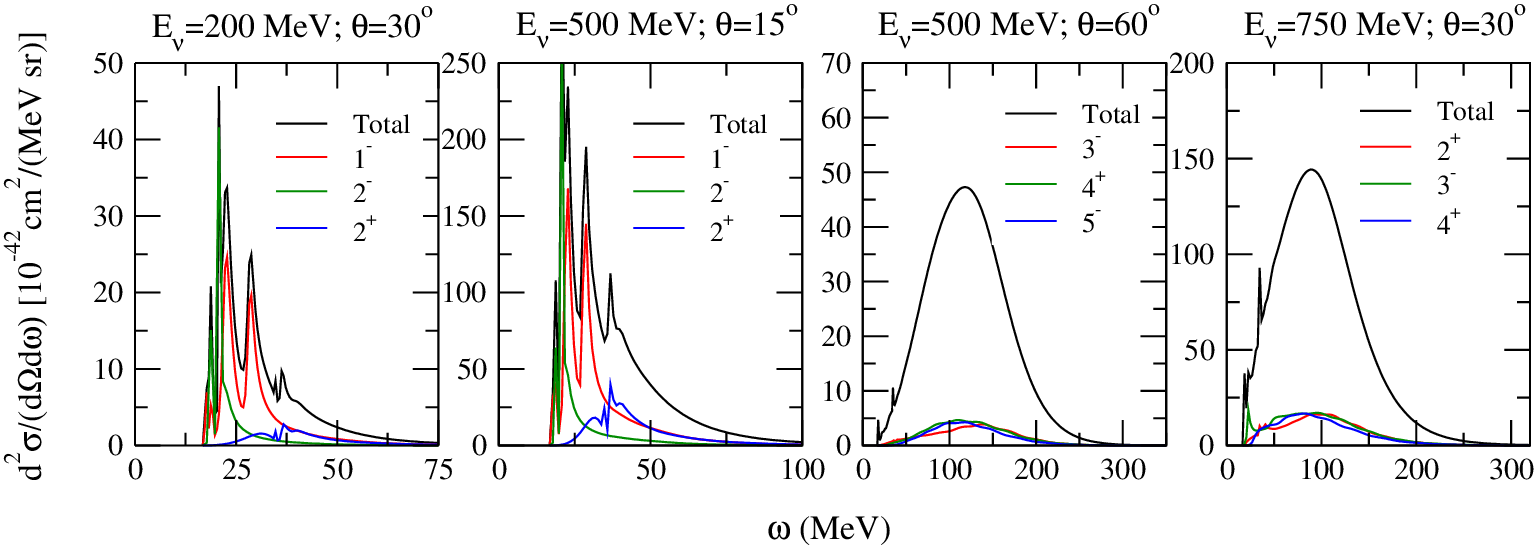}
\caption{(Color online) The three major multipole contributions to the CRPA cross sections in the different kinematical conditions.}
\label{fig_multipoles}
\end{center}
\end{figure}

Another interesting information concerns the main multipoles contributing to the CRPA cross sections, in particular for the resonant structures. To illustrate this point we plot in Fig.~\ref{fig_multipoles} the three major multipole contributions to the cross sections in the four kinematical conditions of Figs.~\ref{fig_LFGRPA_vsMFCRPA} and \ref{fig_cfr_d2s_200_500_750_RPA_only}. As it clearly appears, at low neutrino energy and forward scattering angles, the cross section, with its giant resonance structures is dominated by the $1^-,2^-$ and $2^+$ multipole contributions. On the contrary at larger energies and scattering angles the contribution of the various multipoles is more evenly distributed, as discussed in Ref. \cite{Jachowicz:2002rr}. One can appreciate this point by noticing that the first three major multipole contributions are of the same order of magnitude and are not dominant, hence one needs to include all multipole contributions up to $J=16$ with natural and unnatural parity.   

\begin{figure}
\begin{center}
  \includegraphics[width=9cm,height=6cm]{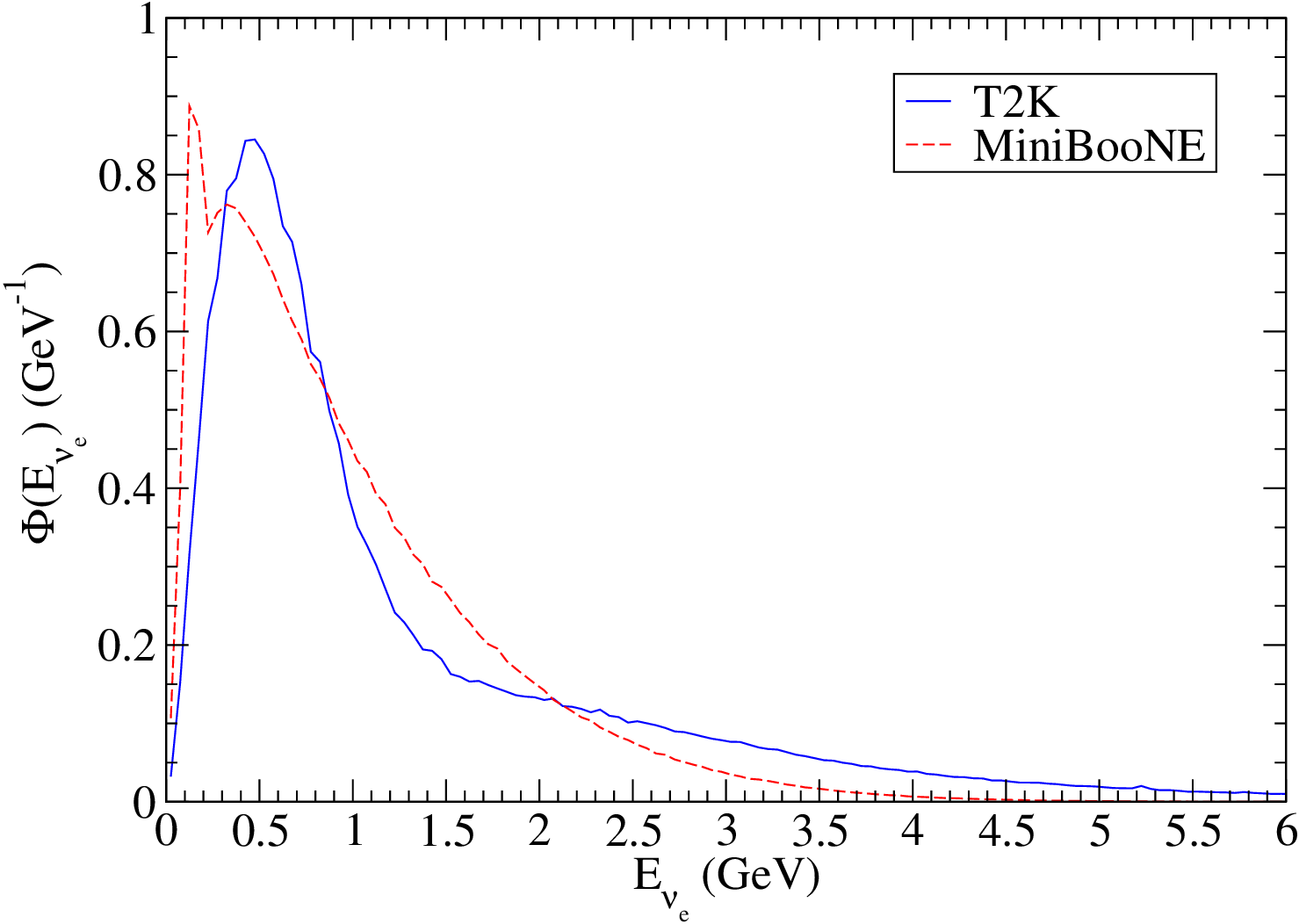}
\caption{(Color online) Normalized electron-neutrino T2K \cite{Abe:2012av}~and MiniBooNE \cite{AguilarArevalo:2008yp}~fluxes.}
\label{fig_flux_nue_t2k_MB}
\end{center}
\end{figure}

\begin{figure}
\begin{center}
  \includegraphics[width=15cm,height=10cm]{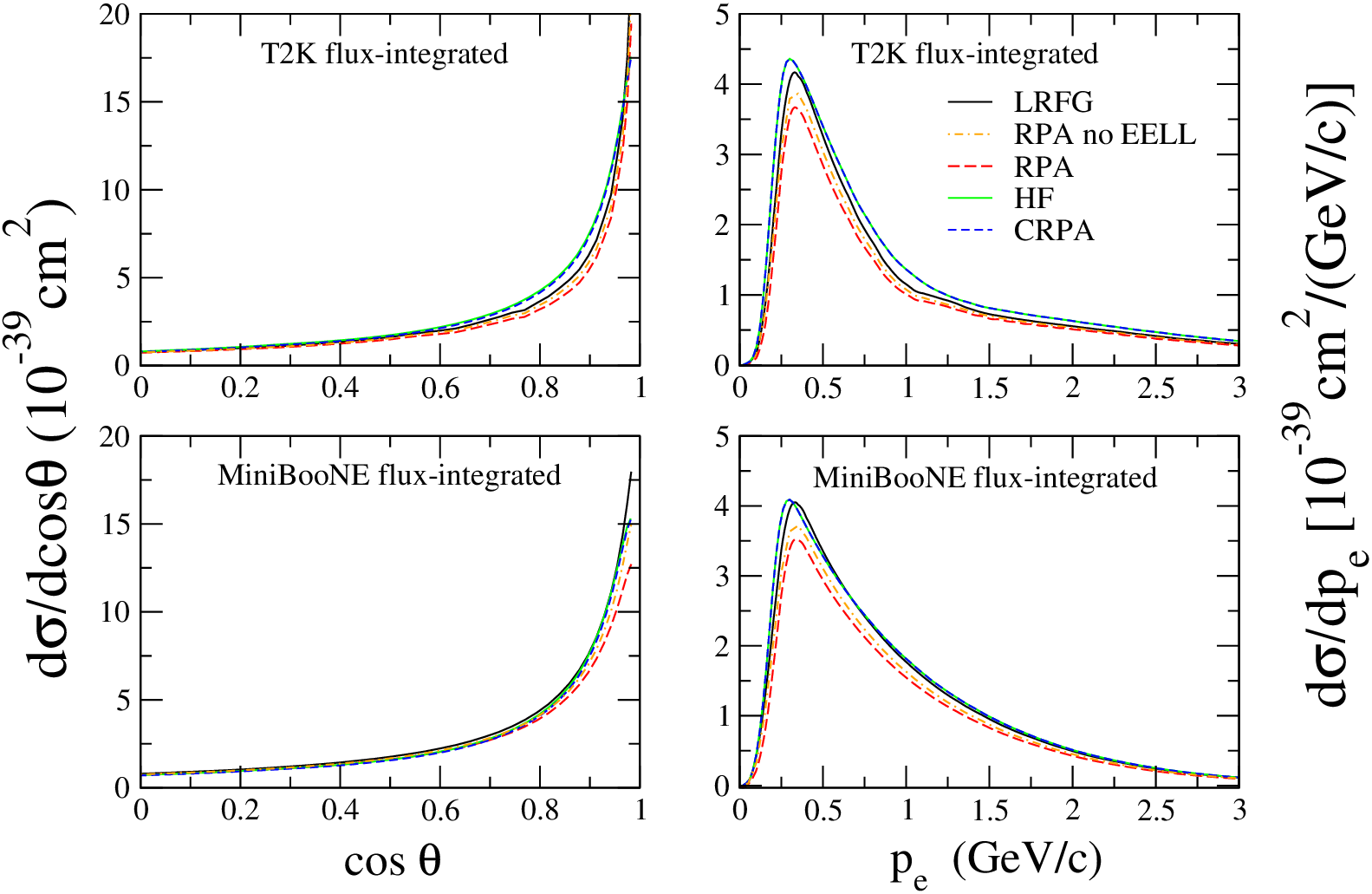}
\caption{(Color online) Electron-neutrino T2K and MiniBooNE flux-folded CC single differential cross sections on Carbon per nucleon.}
\label{fig_ds_fold_mb_t2k_rpa_crpa}
\end{center}
\end{figure}

Turning to the flux folded  cross sections we consider the T2K \cite{Abe:2012av} and MiniBooNE \cite{AguilarArevalo:2008yp} $\nu_e$ normalized fluxes, 
 which are shown in Fig.~\ref{fig_flux_nue_t2k_MB}. We discuss single-differential quasielastic cross sections, $\frac{d \sigma}{d p_e}$ and $\frac{d \sigma}{d \cos \theta_e}$, their theoretical evaluation is displayed in Fig.~\ref{fig_ds_fold_mb_t2k_rpa_crpa}. One observes that the giant resonance effects  are no longer apparent in the CRPA and that in general the differences between HF and CRPA calculations are largely washed out by the flux folding. 
Moreover in the case of MiniBooNE fluxes the HF or CRPA results are very similar to the LRFG ones  while the RPA curves, which are  somewhat below, display  the usual EELL quenching. The results without this EELL quenching are also shown in Fig.~\ref{fig_ds_fold_mb_t2k_rpa_crpa}.  
In the T2K case instead some small differences appear: the HF and CRPA results are above the corresponding LRFG cross section. This difference, which did not show with the MiniBooNE flux, is  the effect of the larger 
T2K high energy tail.  The differences between the two different theoretical models are weighted in different ways by the different flux profiles. But in general the differences between the various models are small and beyond the present level of experimental accuracy.

\subsection{$\nu_e$ \textit{vs} $\nu_\mu$ cross sections}

\begin{figure}
\begin{center}
\includegraphics[width=14.5cm,height=5cm]{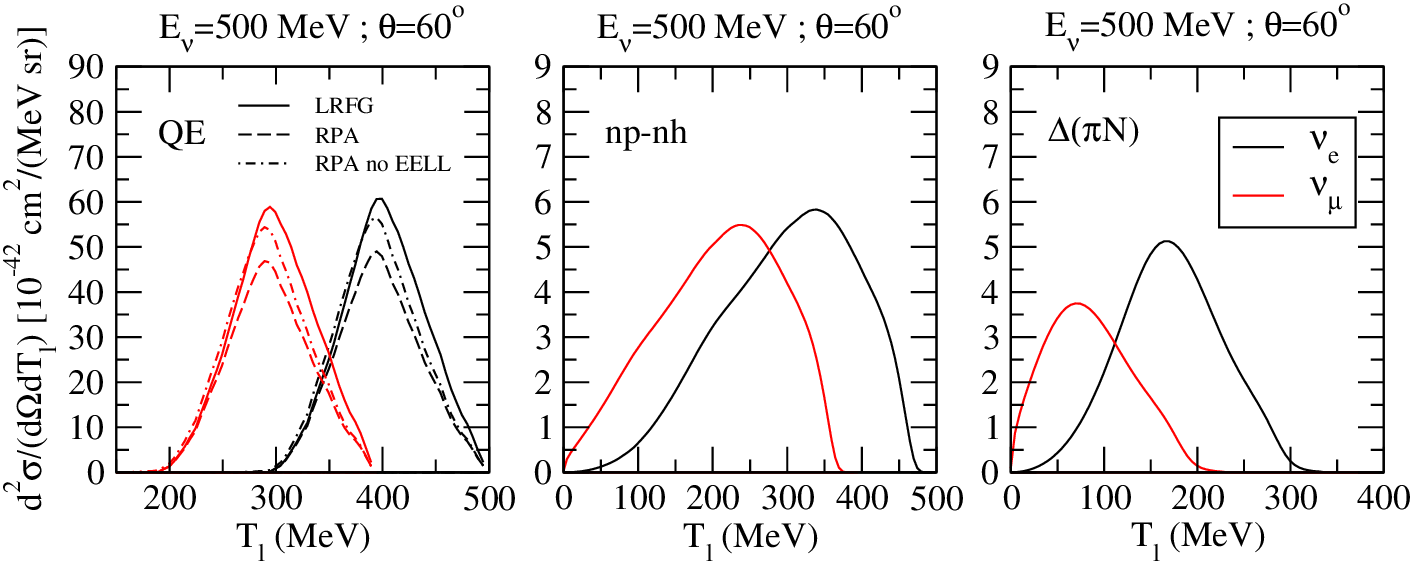}
\caption{(Color online) Electron- and muon-neutrino CC double differential cross section on Carbon for fixed values of scattering angle and 
incident neutrino energie as a function of the charged lepton kinetic energy calculated in the RPA approach of Ref. \cite{Martini:2009uj}. 
Left panel: genuine quasielastic with and without RPA and EELL effects; middle panel: np-nh excitations; right panel: incoherent one pion production contribution.}
\label{fig_numu_nue_vs_Tlepton_3channel.eps}
\end{center}
\end{figure}

\begin{figure}
\begin{center}
 \includegraphics[width=15cm,height=10cm]{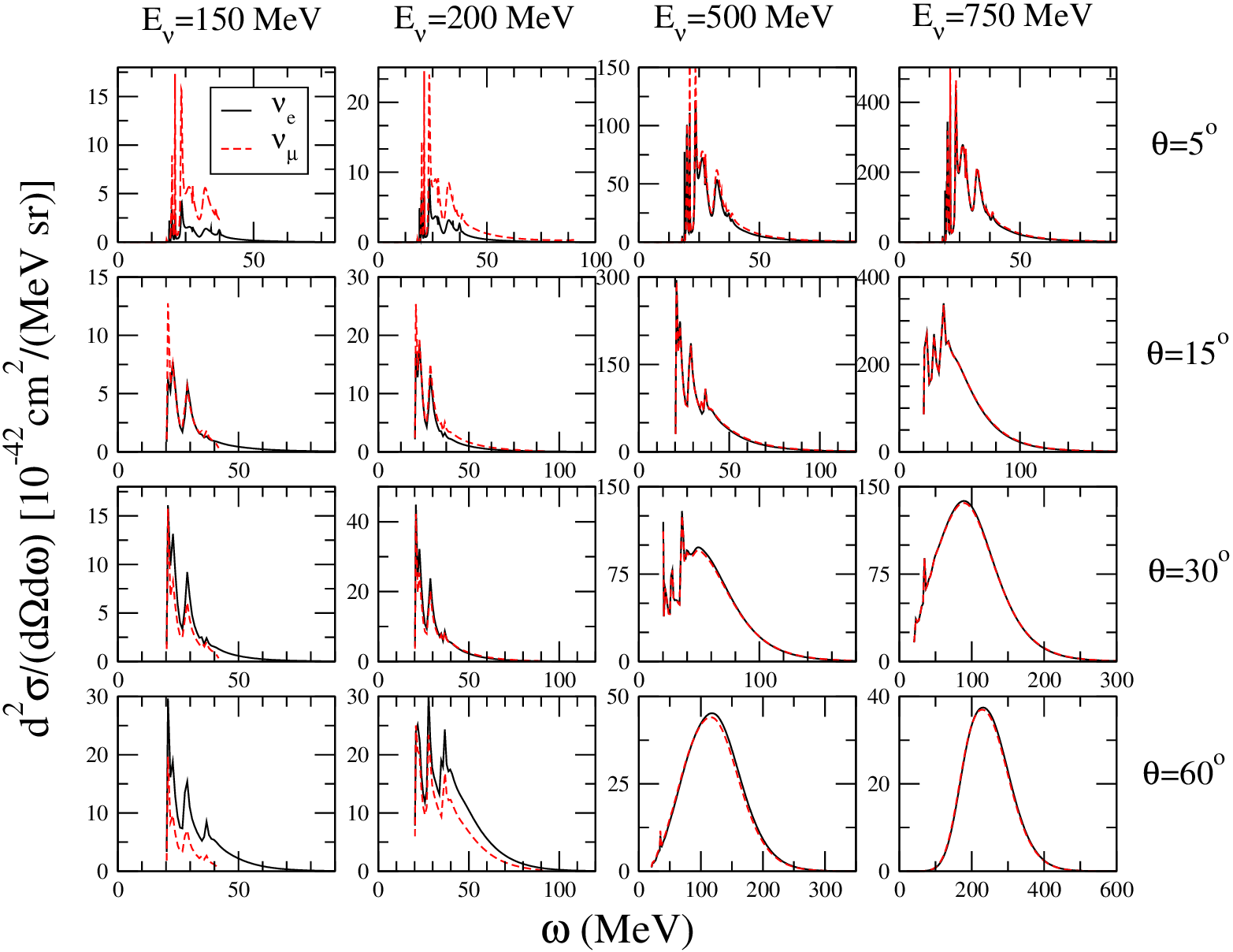}
\caption{(Color online) Electron- and muon-neutrino CC double differential cross section on Carbon calculated in the CRPA approach for fixed values of scattering angles and incident neutrino energies as a function of the energy transferred to the nucleus.}
\label{fig_crpa_only_nue_numu_15_30_60_100_750}
\end{center}
\end{figure}

\begin{figure}
\begin{center}
 \includegraphics[width=15cm,height=10cm]{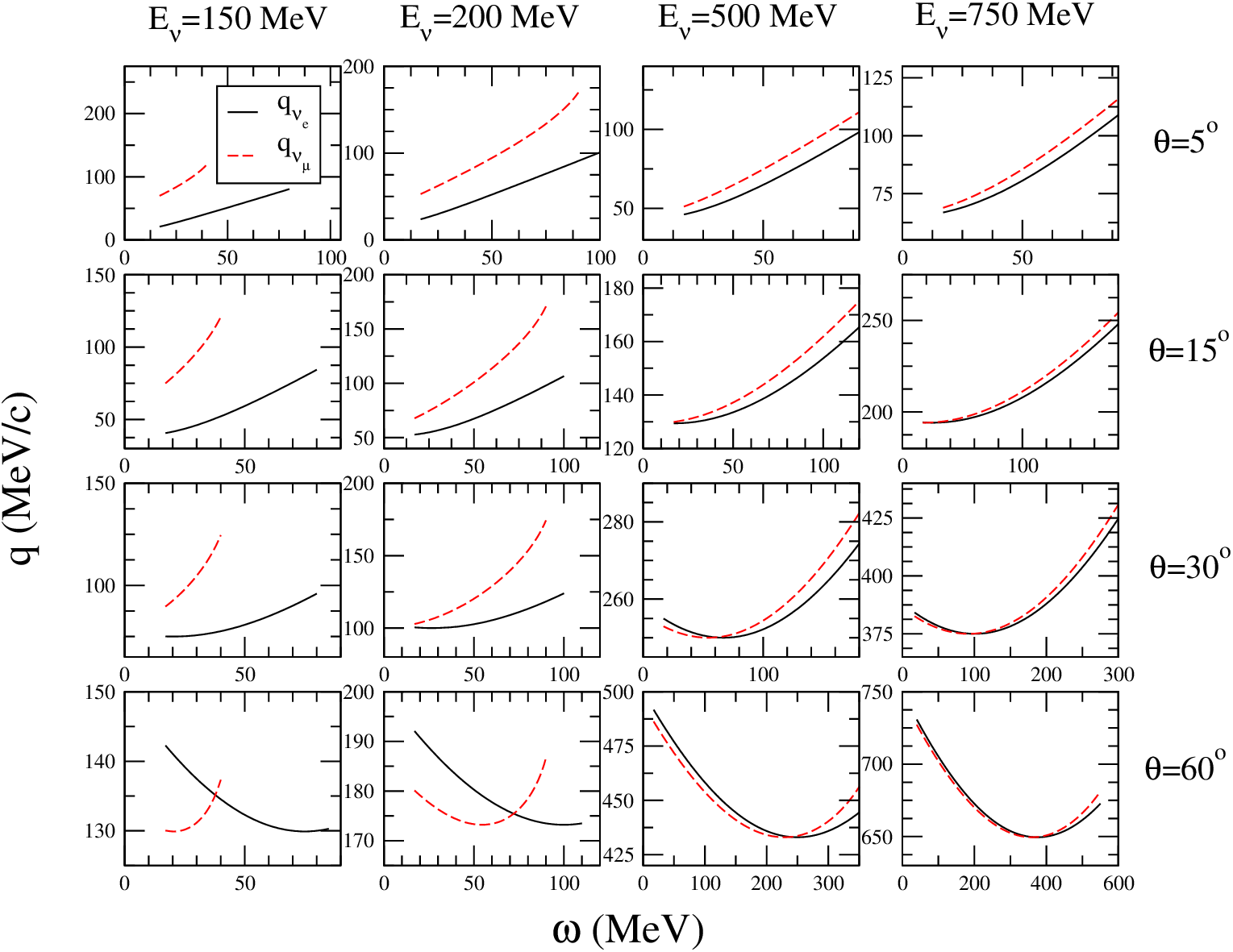}
\caption{(Color online) The values of the momentum transfer versus the transferred energy for the $\nu_e$ and $\nu_\mu$ scattering corresponding to the kinematical conditions of Fig.~\ref{fig_crpa_only_nue_numu_15_30_60_100_750}.}
\label{fig_qmu_qe}
\end{center}
\end{figure}

\begin{figure}
\begin{center}
\includegraphics[width=12cm,height=8cm]{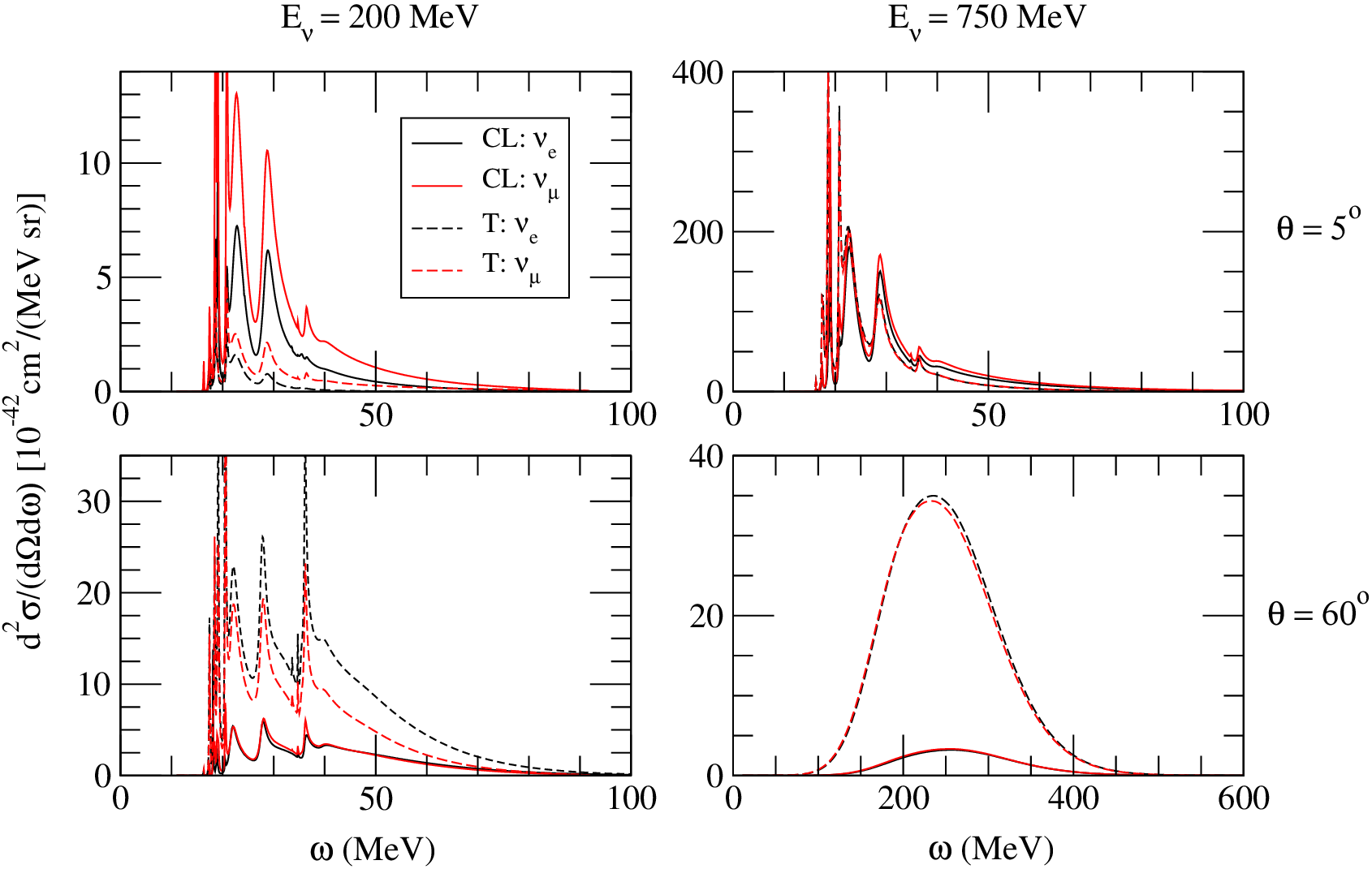}
\caption{(Color online) Coulomb-longitudinal (CL) and transverse (T) contributions to electron- and muon-neutrino CC double differential cross section on Carbon calculated in the CRPA approach for incident neutrino energies of 200 MeV and 750 MeV and two fixed values of scattering angles as a function of the transferred energy to the nucleus.}
\label{fig_nue_numu_crpa_cl_t_E200_final}
\end{center}
\end{figure}

\begin{figure}
\begin{center}
  \includegraphics[width=15cm,height=10cm]{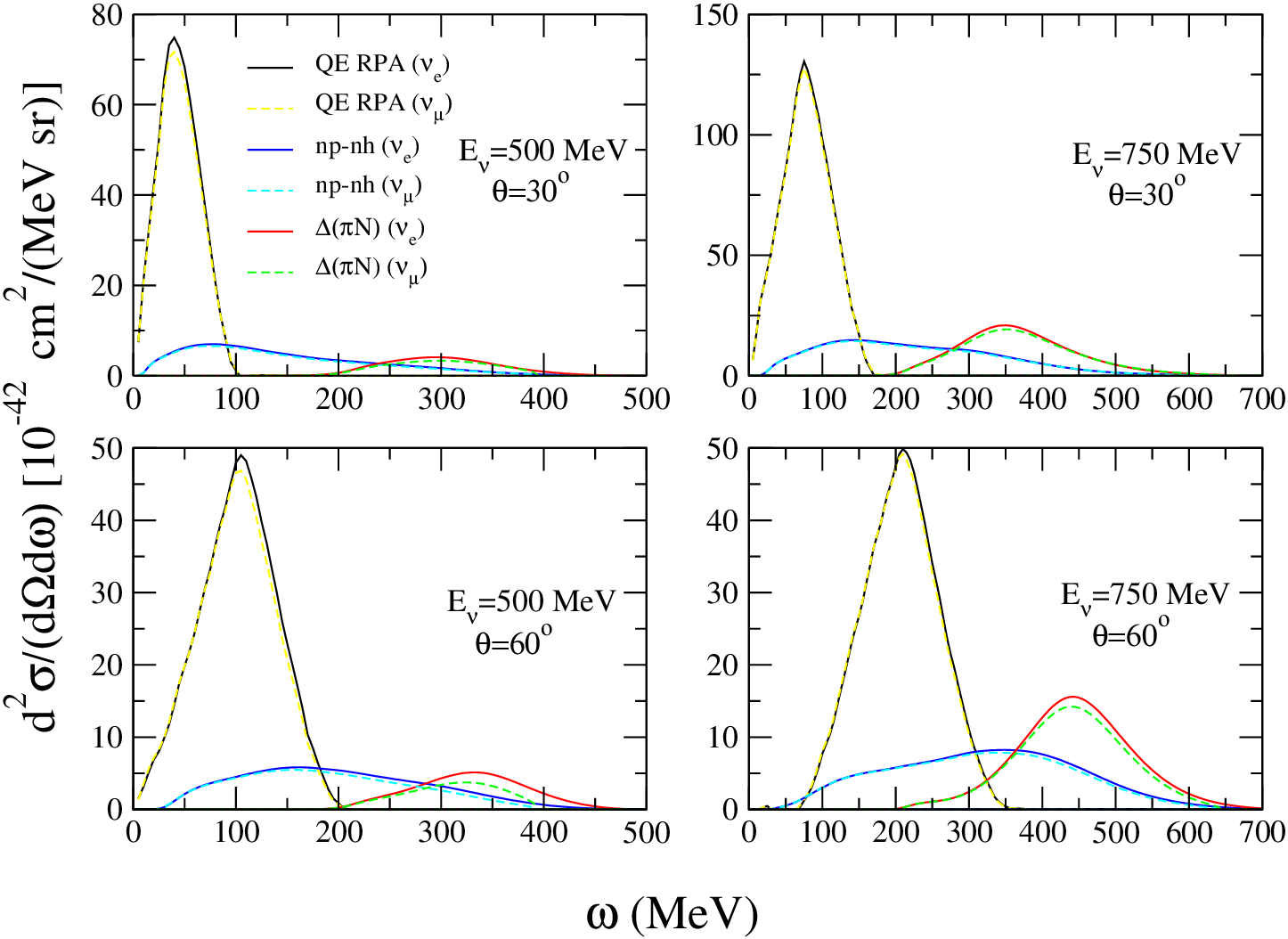}
\caption{(Color online) Electron- and muon-neutrino CC double differential cross section on Carbon calculated in the RPA approach of Martini \textit{et al.} for fixed values of scattering angles and incident neutrino energies as a function of the transferred energy to the nucleus. The genuine quasielastic (QE), multinucleon (np-nh), and incoherent 
one-pion production excitations are plotted separately.}
\label{fig_30_60_500_750_qe_np_pi_nue_numu}
\end{center}
\end{figure}

After the discussion of the differences between the two approaches for the $\nu_e$ case, 
we turn to a comparison between the charged current $\nu_e$ and $\nu_\mu$ cross sections. We perform this comparison for different final state channels.   
For the sake of illustration, we present in Fig.~\ref{fig_numu_nue_vs_Tlepton_3channel.eps} the double differential cross sections 
in different channels (genuine quasielastic, multinucleon excitations and one pion production) 
for fixed values of the scattering angle (60 degrees) and neutrino energy ($E_\nu$=500 MeV) as a function of the lepton kinetic energy $T_l$, a measurable quantity. 
The role of the different charged lepton masses appears not only in the trivial relative shift between the $\nu_e$ and $\nu_\mu$ CC cross sections, 
according to the identities
\begin{equation}
T_l=E_l-m_{\textrm{lepton}}=E_\nu-\omega-m_{\textrm{lepton}}=\omega_{max}-\omega,
\end{equation}
but also in the strength and in the shape of the cross sections, particularly in the pion production channel. 
In order to eliminate trivial mass shift effects we plot in the following figures (Figs. \ref{fig_crpa_only_nue_numu_15_30_60_100_750},  \ref{fig_nue_numu_crpa_cl_t_E200_final} and \ref{fig_30_60_500_750_qe_np_pi_nue_numu}) the differential cross sections as a function of the energy transfer $\omega = E_\nu-m_{\textrm{lepton}}- T_l$. 

We start with the CRPA case which allows a simultaneous treatment of giant resonances and quasielastic excitations. 
In Fig.~\ref{fig_crpa_only_nue_numu_15_30_60_100_750}, we display the double-differential cross sections for different values of incoming neutrino energy and lepton scattering angle, both for $\nu_e$ and $\nu_\mu$. In most cases the $\nu_e$ and $\nu_\mu$ results are quite similar, sometimes practically indistinguishable. However, in some cases interesting differences appear. 
The first one is a consequence of the stringent limit on the maximum transferred energy 
$\omega_{max}=E_\nu-m_{\textrm{lepton}}$ which has smaller values in the muon case. This threshold effect can be observed in Fig.~\ref{fig_crpa_only_nue_numu_15_30_60_100_750} for $E_\nu$=150 MeV and for $E_\nu$=200 MeV in the case of 60 degrees. 
Other differences can be appreciated by observing the evolution with the scattering angle of the cross sections at small neutrino energies 
such as $E_\nu$=150 MeV or $E_\nu$=200 MeV. For small scattering angles such as 5 degrees, $\nu_\mu$ cross sections are higher than the $\nu_e$ ones, while for larger scattering angles, for example 60 degrees this behaviour is opposite. 
At intermediate angles the two cross sections are closer to each other. 
This angular behavior weakly survives at $E_\nu$=500 MeV while for $E_\nu$=750 MeV the $\nu_e$ and $\nu_\mu$ cross sections practically coincide for all the scattering angles. 
The differences discussed above are also related to the differences in the momentum transfer between the $\nu_e$ and $\nu_\mu$ scattering. 
For completeness we show in Fig.~\ref{fig_qmu_qe} the values of the momentum transfer spanned in the 16 panels of Fig.~\ref{fig_crpa_only_nue_numu_15_30_60_100_750}. As expected, the major differences between $\nu_e$ and $\nu_\mu$ appear at small neutrino energies 
where threshold effects are more evident. Furthermore at small scattering angle the momentum transfer is always larger 
for $\nu_\mu$ than for $\nu_e$ while at 60 degrees this is not always the case.

It is also interesting to illustrate the behavior of $\nu_e$ and $\nu_\mu$ cross sections by separating their contributions, 
as shown in Fig.\ref{fig_nue_numu_crpa_cl_t_E200_final} for incoming neutrino energies of $E_\nu$=200 MeV and $E_\nu$=750 MeV. 
According to the notation of Ref. \cite{Pandey:2013cca}, the global contribution related to the Coulomb and longitudinal multipole excitation operators (containing vector and axial components) is labeled as 
CL. In the language of Refs.~\cite{Martini:2009uj,Martini:2010ex} it represents the sum of isovector and isospin spin-longitudinal response contributions.  
The sum of transverse contributions, including the vector-axial interference term, is labeled as T. 
These are the terms containing the isospin spin-transverse response in the language of Refs.~\cite{Martini:2009uj,Martini:2010ex}.
As one can observe in Fig.\ref{fig_nue_numu_crpa_cl_t_E200_final}, for $E_\nu$=200 MeV and $\theta$=5 degrees (and in general for very forward scattering) 
the neutrino cross section is dominated by the CL contribution while for larger angles, such as 60 degrees, the transverse contribution T is dominant. 
At larger energies the transverse part dominates everywhere except for very small scattering angles. 
At $E_\nu$=200 MeV and $\theta=5$ degrees the dominant CL contribution to the
cross sections, as well as the smaller T one, are larger for $\nu_\mu$ than for $\nu_e$, hence the larger $\nu_\mu$ cross sections for this case. The relative weight of CL and T contributions is the result of a subtle interplay between lepton kinematic factors and response functions. 
The competition for dominance of the cross section between both, is very sensitive to energy and momentum transfer. The surprising dominance of $\nu_\mu$ over $\nu_e$ cross sections for small scattering angles is related to this and dictated by the non-trivial dependence of momentum transfer on lepton mass and scattering angle for forward scattering, as illustrated in Fig.~\ref{fig_qmu_qe}.

\begin{figure}
\begin{center}
  \includegraphics[width=12cm,height=8cm]{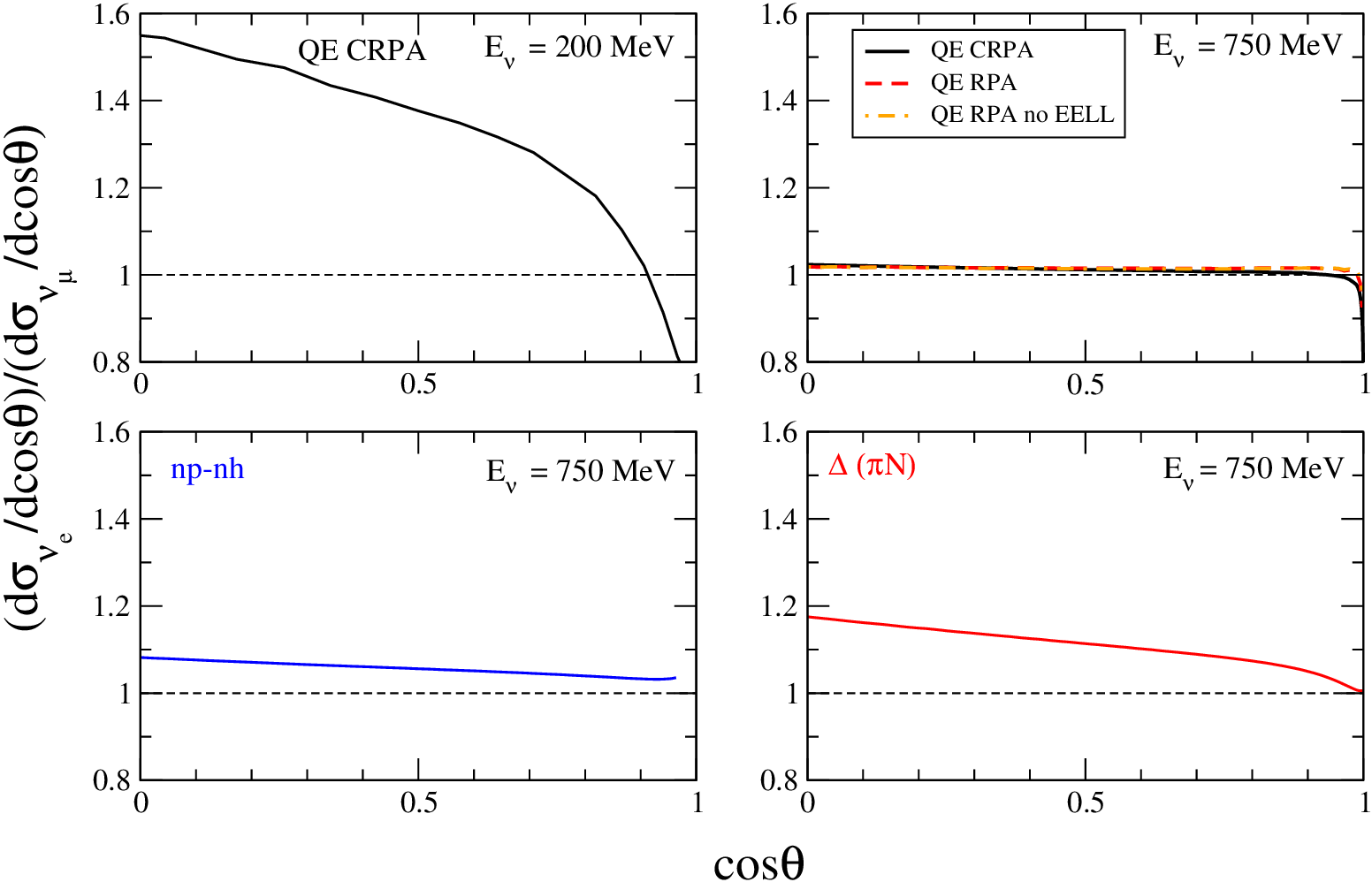}
\caption{(Color online) Ratio of the $\nu_e$ over $\nu_\mu$  differential cross section on Carbon calculated for two fixed values of incident neutrino energies as a function of the cosine of the lepton scattering angle. The 1p-1h results in the CRPA approach are shown for $E_\nu$=200 MeV and $E_\nu$=750 MeV. 
The 1p-1h results in the RPA approach with or without the EELL effect, the np-nh excitations and the one pion production (via $\Delta$ excitation) results are shown for $E_\nu$=750 MeV.}
\label{fig_nue_numu_crpa_ratio_E200_final}
\end{center}
\end{figure}

The non-trivial behaviour of the $\nu_e$ cross sections with respect to the $\nu_\mu$ ones 
is also illustrated 
in Fig.~\ref{fig_nue_numu_crpa_ratio_E200_final} where the ratio of the single differential cross section 
$\frac{d \sigma_{\nu_e}}{d \cos \theta}/\frac{d \sigma_{\nu_\mu}} {d \cos \theta}$ 
is shown for $E_\nu$=200 MeV and $E_\nu$=750 MeV. 
In the low energy case of $E_\nu$=200 MeV where the semiclassical description breaks down, we restrict ourself to the CRPA approach.  
In this case the 
ratio deviates very appreciably from 1 while at 
larger neutrino energies, such as $E_\nu$=750 MeV, it gets closer to 1 in the CRPA as well as in the RPA case with or without the EELL effect. In Fig.~\ref{fig_nue_numu_crpa_ratio_E200_final} this quantity $\frac{d \sigma_{\nu_e}}{d \cos \theta}/\frac{d \sigma_{\nu_\mu}} {d \cos \theta}$ at $E_\nu$=750 MeV is given also for two other channels, the pion production and multinucleon excitations in the RPA approach of Martini \textit{et al.} where they are available. 
This $\frac{d \sigma_{\nu_e}}{d \cos \theta}/\frac{d \sigma_{\nu_\mu}} {d \cos \theta}$ ratio, always larger than 1, is characterized by a smooth decreasing behavior. 
For the pion emission channel (via $\Delta$ excitation) this ratio is larger than the one for the np-nh and 1p-1h excitations. 

Concerning the pion production and multinucleon excitations, we display for completeness in Fig.~\ref{fig_30_60_500_750_qe_np_pi_nue_numu} the $\nu_e$ and $\nu_\mu$ results obtained for these channels 
(as well as for the QE one) in the RPA approach of Martini \textit{et al.} 
for the double differential cross sections at incoming neutrino energies of $E_{\nu}=500$ MeV and $E_{\nu}=750$ MeV
and scattering angles of 30 and 60 degrees. 
One observes the clear energy separation between the three channels, the highest energy transfer occurring for pion emission. 
Ignoring Fermi momentum and RPA reshaping effects, the quasi elastic peak occurs for an energy transfer $\omega= Q^2/(2M_N)$ where 
$Q^2 = q^2-{\omega}^2=2 E_\nu E_l (1-\cos\theta)- m_l^2 +2 E_\nu(E_l- P_l)\cos \theta$. In the electron case where $m_l=0$ it leads to 
$\omega={E_\nu}^2 (1-\cos\theta)/(M_N + E_\nu (1-\cos\theta))$. 
As for pion emission, in our model  it occurs via $\Delta$ excitation. 
In the same (nucleons at rest) approximation the pion emission peak is shifted towards large energy transfer, with the condition $ \omega = Q^2/(2M_N) +  \Delta  M$  with  
$ \Delta  M =(M^2_{\Delta}-M^2_N)/2M_N= 338$ MeV. 
For $\nu_e$ this leads to $\omega=( M_N{ \Delta  M +E_\nu}^2 (1-\cos\theta))/(M_N + E_\nu (1-\cos\theta))$. 
These formulas explain the positions of the quasielastic and $\Delta$ peaks. 
As for the multinucleon excitations they lie between the two. 
The difference between the $\nu_e$ and $\nu_\mu$ cross sections mostly shows up in the energy transfer limit which is $\omega_{max}\simeq E_\nu$ for electrons and $\omega_{max}=E_\nu- m_{\mu}$ for muons. 
Hence it shows up mostly for pion production and it is more pronounced at low neutrino energies. It is also more pronounced at large scattering angles 
since the double differential cross sections move towards larger $\omega$ when the scattering angle increases. 
This behavior with the scattering angle appears also in the previous Fig.~\ref{fig_nue_numu_crpa_ratio_E200_final}.

\section{Comparison with the T2K $\nu_e$ inclusive cross sections} 
\label{sec_comparison_exp}
\begin{figure}
\begin{center}
  \includegraphics[width=12cm,height=8cm]{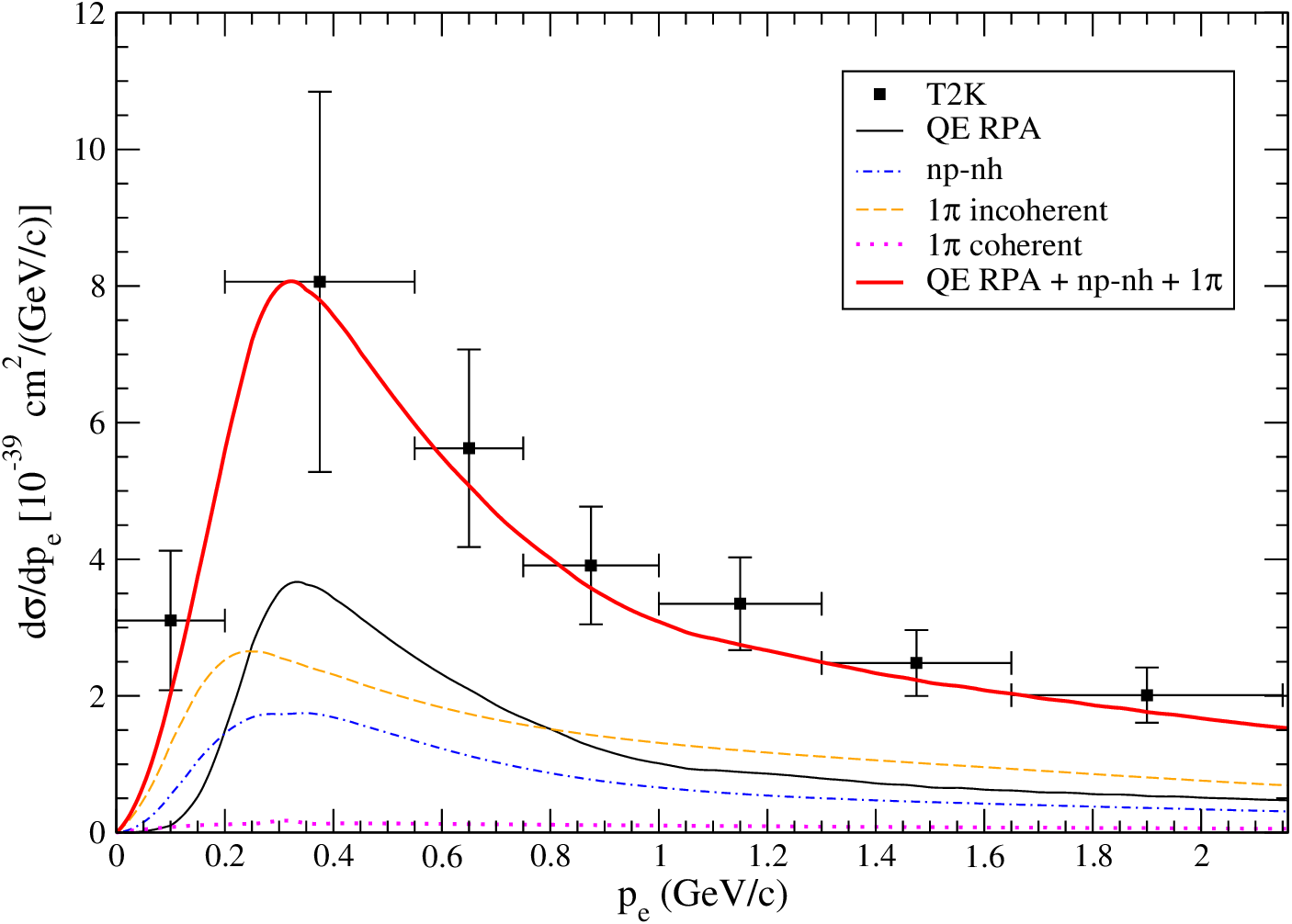}
\caption{(color online). T2K flux-integrated inclusive 
$\nu_e$ CC differential cross section on Carbon per nucleon
as a function of the electron momentum. The different contributions to this inclusive cross section 
obtained in the Martini \textit{et al.} RPA model of Ref. \cite{Martini:2009uj} are shown.   
The experimental T2K points are taken from Ref. \cite{Abe:2014agb}.}
\label{fig_t2k_ds_dpe}
\end{center}
\end{figure}
\begin{figure}
\begin{center}
  \includegraphics[width=12cm,height=8cm]{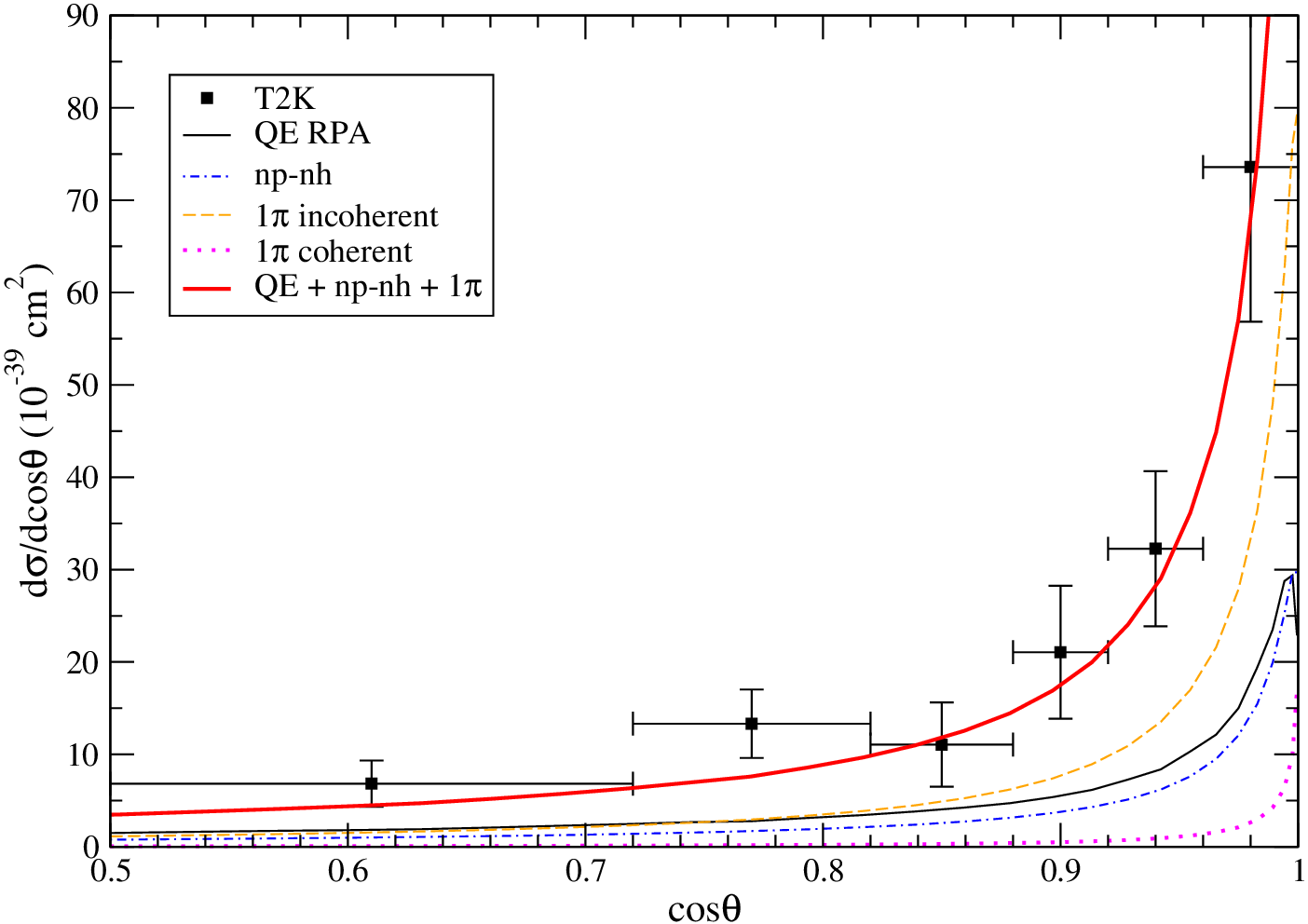}
\caption{(color online). T2K flux-integrated inclusive 
$\nu_e$ CC differential cross section on Carbon per nucleon
as a function of the cosine of the lepton scattering angle. The different contributions to this inclusive cross section obtained in the Martini \textit{et al.} RPA 
model of Ref. \cite{Martini:2009uj} are shown.   
The experimental T2K points are taken from Ref. \cite{Abe:2014agb}.}
\label{fig_t2k_ds_dcos_e}
\end{center}
\end{figure}

The T2K collaboration published the first results for $\nu_e$ charged-current inclusive differential cross sections on Carbon \cite{Abe:2014agb}. 
In this section we compare these experimental results with our predictions, restricting to the Martini \textit{et al.} \cite{Martini:2009uj} RPA approach which describes all final channels. 
We compute the $\nu_e$ T2K flux integrated differential cross sections 
$\frac{d \sigma}{d p_e}$ and $\frac{d \sigma}{d \cos \theta_e}$ in the different excitation channels, namely quasielastic, 
multinucleon excitations (np-nh) and one-pion (coherent and incoherent) production. 
The $\nu_e$ T2K flux averaging favors large $E_\nu$ values which ensures the validity of the semi-classical RPA approach of Martini \textit{et al.} 
In Figs.~\ref{fig_t2k_ds_dpe} and \ref{fig_t2k_ds_dcos_e} we plot the different exclusive channel contributions separately, 
as well as their sum. This sum is in a good agreement with the experiment. 
Notice that this agreement needs the presence of the np-nh contribution 
(which even dominates the genuine QE one for small $p_e$ values, $p_e \lesssim$ 0.2 GeV), a conclusion already reached by Martini and Ericson \cite{Martini:2014dqa} in connection with the T2K inclusive $\nu_\mu$ double differential cross sections \cite{Abe:2013jth}. 
This agreement with both $\nu_\mu$ and $\nu_e$ CC inclusive T2K flux folded differential cross sections is not systematically 
obtained in other approaches. For instance the SuSAv2 model by Ivanov
\textit{et al.} \cite{Ivanov:2015aya} reproduces  well the CC inclusive T2K flux folded $\nu_\mu$ double differential cross section but underestimates the CC inclusive T2K flux folded $\nu_e$ single differential cross section. A comparison with these quantities has also been performed by Meucci and Giusti using the Relativistic Green's function model which turned to underestimate the $\nu_\mu$ and $\nu_e$ CC inclusive T2K data \cite{Meucci:2015bea}. 

\section{Summary and Conclusions}
In conclusion, our study has dealt with several facets of the neutrino interaction with nuclei. 
A large part is devoted to the comparison between two different approaches   
to describe the interaction of neutrinos with nuclei. Both go beyond the 
impulse approximation and take into account, albeit
in different ways, the interaction between nucleons. 
The CRPA approach of Jachowicz \textit{et al.} starts from a continuum Hartree Fock description with Skyrme type interactions. 
The shell structure of the nucleus is present in this approach.  
The RPA-based approach of Martini \textit{et al.} 
instead starts from a semiclassical description of the bare polarization propagator with a realistic nuclear density distribution. 
The shell structure is ignored in this description. The RPA effects also differ in the two approaches. For the residual interaction the first method uses the same Skyrme interaction as for the mean field, while in the approach of Martini \textit{et al.}, it is parametrized in terms of pion and  rho exchange and a contact Landau Migdal interaction. But the main difference is the possibility of mixing of $\Delta$-hole states in the second approach.  It produces a general quenching of the responses which shows up in most kinematical  conditions that we have explored. The CRPA of Jachowicz \textit{et al.} allows a description of giant resonances and quasielastic excitations while the RPA evaluations of Martini \textit{et al.} includes quasielastic  but also coherent and incoherent pion production, and multinucleon excitations. 

We have compared the two approaches for the 
one nucleon - one hole excitations finding a reasonable agreement between them in the quasielastic peak region, 
with a trend for the RPA approach to lead to lower cross sections than the CRPA presumably due to the mixing with $\Delta$ excitations. 
Other general trends are related to the more important high transferred-energy tail in the CRPA results and to a relative shift 
of the cross sections of $\omega\simeq 18$ MeV, reflecting the presence of the nucleon separation energy in the CRPA calculations.   
The most striking difference is the appearance of giant resonance peaks in the CRPA results. 
The comparison of the two approaches has been performed for fixed values of the incoming neutrino energy as well as for the $\nu_e$ T2K and MiniBooNE flux-folded cross sections. 

We have also compared the $\nu_e$ cross sections with the corresponding $\nu_\mu$ ones for fixed values of the neutrino energy in order to investigate the 
impact of different charged lepton masses. We have found some non trivial behaviour, 
in particular for the 1p-1h excitations at low neutrino energies, 
such as an inversion with the scattering angle of the relative strength of $\nu_e$ and $\nu_\mu$ cross sections. 
Due to the different kinematical limits, the $\nu_e$ cross sections are in general expected to be larger than the $\nu_\mu$ ones, 
 however for forward scattering angles this hierarchy is opposite. In the precision era of neutrino oscillation physics the $\nu_e$ cross sections should be known with the same accuracy as the $\nu_\mu$ ones. 
Trying to deduce the $\nu_e$ cross sections from the experimental $\nu_\mu$ ones can be considered only as a first approximation in the study of the $\nu_e$ interactions.

Concerning the comparison with experiment, we have considered the inclusive $\nu_e$ T2K flux-folded single-differential cross sections on Carbon 
where the inclusion of np-nh and pion production effects beyond the genuine quasielastic is mandatory. 
For this reason we have compared the data with the only RPA-based approach of Martini \textit{et al.} which includes all these channels. 
We have found a good agreement with the data. 
This success obtained with a new flux, the $\nu_e$ T2K one, complements those already reached with the three different fluxes, such as the MiniBooNE $\nu_\mu$ and $\bar{\nu}_\mu$, and T2K $\nu_\mu$ ones.

\begin{acknowledgements}
This work was supported by the Interuniversity Attraction Poles Programme initiated by the Belgian Science Policy Office (BriX network P7/12) and by the Research
Foundation Flanders (FWO-Flanders). M.M. acknowledges also the support and the framework of the ``Espace de Structure et de r\'eactions Nucl\'eaire Th\'eorique'' (ESNT, \url{http://esnt.cea.fr} ) at CEA.
We thank Ra\'ul Gonz\'alez-Jim\'enez, Carlotta Giusti and Andrea Meucci for useful discussions. 
\end{acknowledgements}

\end{document}